\documentclass[twocolumn]{aastex631}
\UseRawInputEncoding
\usepackage{amsmath}
\usepackage{amssymb}
\usepackage{mathtools}
\usepackage{diagbox}
\usepackage{xcolor}
\usepackage{pgf}
\usepackage{collcell}

\shorttitle{HCI}
\shortauthors{D. Gazith B. Zackay}

\graphicspath{{./}{}}

\begin{document}

\title{Precision speckle pattern reconstruction for high contrast imaging}

\author{Dotan Gazith}

\author{Barak Zackay}
\affiliation{Dept. of Particle Phys. \& Astrophys., Weizmann Institute of Science, Rehovot 76100, Israel}

\begin{abstract}

In High Contrast Imaging, a large instrumental, technological and algorithmic effort is made to reduce residual speckle noise and improve the detection capabilities. 
In this work, we explore the potential of using a precise physical description of speckle images, in conjunction with the optimal detection statistic to perform High Contrast Imaging.
Our method uses short-exposure speckle images, reconstructing the Point Spread Function (PSF) of each image with phase retrieval algorithms. Using the reconstructed PSF's we calculate the optimal detection statistic for all images. We analyze the arising bias due to the use of a reconstructed PSF and correct for it completely up to its accumulation over $10^4$ images. We measure in simulations the method's sensitivity loss due to overfitting in the reconstruction process and get to an estimated 5$\sigma$ detection limit of $5\times 10^{-7}$ flux ratio at angular separations of $0.1 -0.5^{\prime\prime}$ for a $1h$ observation of Sirius A with a 2m-telescope.

\end{abstract}



\section{Introduction} \label{sec:intro}

High Contrast Imaging (HCI), the method to detect exoplanets in which the exoplanet is seen directly as an additional source near its host star, is an important method for current and future characterization of exoplanets atmospheres (\cite{currie_combined_2011}, \cite{konopacky_detection_2013}) as it measures light from the exoplanet itself.

HCI requires separating the faint exoplanet from its bright nearby host star, which is highly dependent on the telescope's angular resolution. For ground-based observatories, the atmosphere presents an additional constraint on the telescope's angular resolution. The turbulent flow of air in the atmosphere creates index of refraction variations, that create rapidly changing phase aberrations. The phase aberrations degrade the telescope's angular resolution, creating speckle images for short-exposure imaging and a broad seeing disk image for long-exposure.

A way to overcome atmospheric seeing is the use of Adaptive Optics (AO) systems that sense and correct, in real-time, the atmospheric phase aberrations. Such systems bring the angular resolution from the seeing limit close to the telescope's diffraction scale and are used by the leading HCI instruments on state-of-the-art telescopes (VLT-SPHERE \cite{beuzit_sphere_2019}, Subaru-CHARIS \cite{groff_charis_2015} and MEC \cite{walter_mkid_2020}, Gemini-GPI \cite{nielsen_gemini_2019}).

The biggest challenge for those AO-based HCI instruments are residual aberrations not corrected or sensed by the AO system \cite{mawet_review_2012}, those aberrations create quasi-static speckles, slowly varying speckles that are confused as point sources.
A wide range of efforts are made to tackle this challenge, better wavefront sensing (eg. \cite{baudoz_self-coherent_2005}, \cite{skaf_high_2021}, \cite{skaf_-sky_2022}), deformable mirror technological improvements (eg. \cite{madec_overview_2012}), nulling coronagraphy (eg. \cite{ruane_review_2018}) and post-processing methods (eg. \cite{marois_angular_2006}, \cite{racine_speckle_1999}, \cite{lafreniere_new_2007}, \cite{rodack_millisecond_2021}, \cite{frazin_millisecond_2021}).

\begin{figure}[htbp]
\plotone{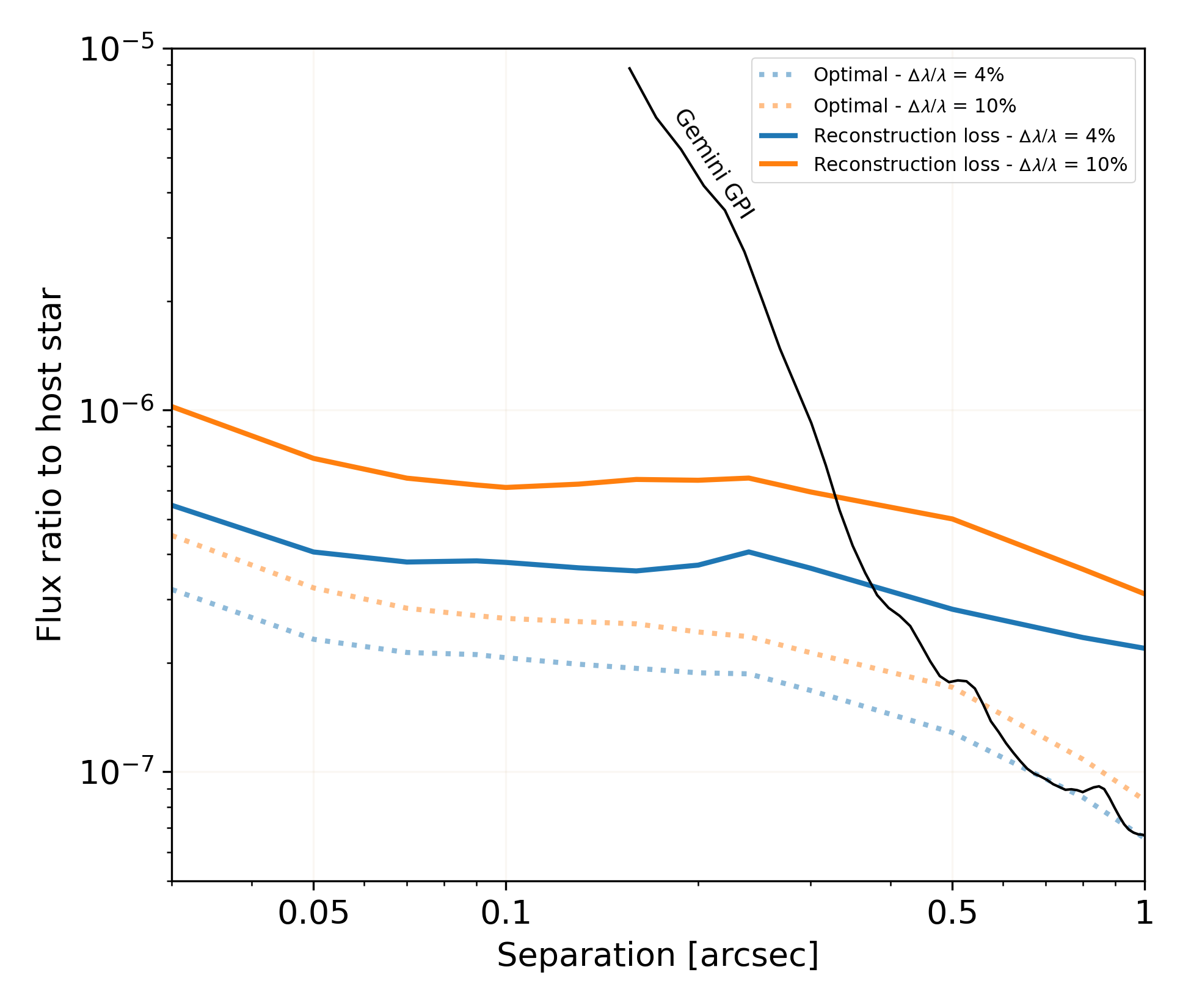}
\caption{$5\sigma$ Detection limit at different separations. The detection limit is calculated for a $2\mathrm{m}$-telescope, at seeing condition $r_0=15\mathrm{cm}$, for total of $10^{14}$ photons (corresponds to observing Sirius A for $\sim1\mathrm{h}$ at 500-700nm with efficiency of $\sim70\%$). The colored-dotted lines are the optimal detection limit described in Section \ref{sec:hci_turbulence}, the colored-solid include overfitting losses described in Section \ref{sec:methods}, the black line shows the performance of GPI from \cite{ruffio_improving_2017}. The plot was created using the package by \cite{bailey_cgi-flux-ratio-plot_2021}.
\label{fig:contrast_curve}}
\end{figure}

Analyzing the idealized case of HCI with perfectly known, yet uncorrected, Point Spread Function (PSF), we obtain that even a modest-sized telescope in the seeing limited case can reach a fantastic contrast, comparable to the best performance of state-of-the-art facilities, as can be seen in the dashed lines in Figure \ref{fig:contrast_curve}.

\begin{figure}[htbp]
\plotone{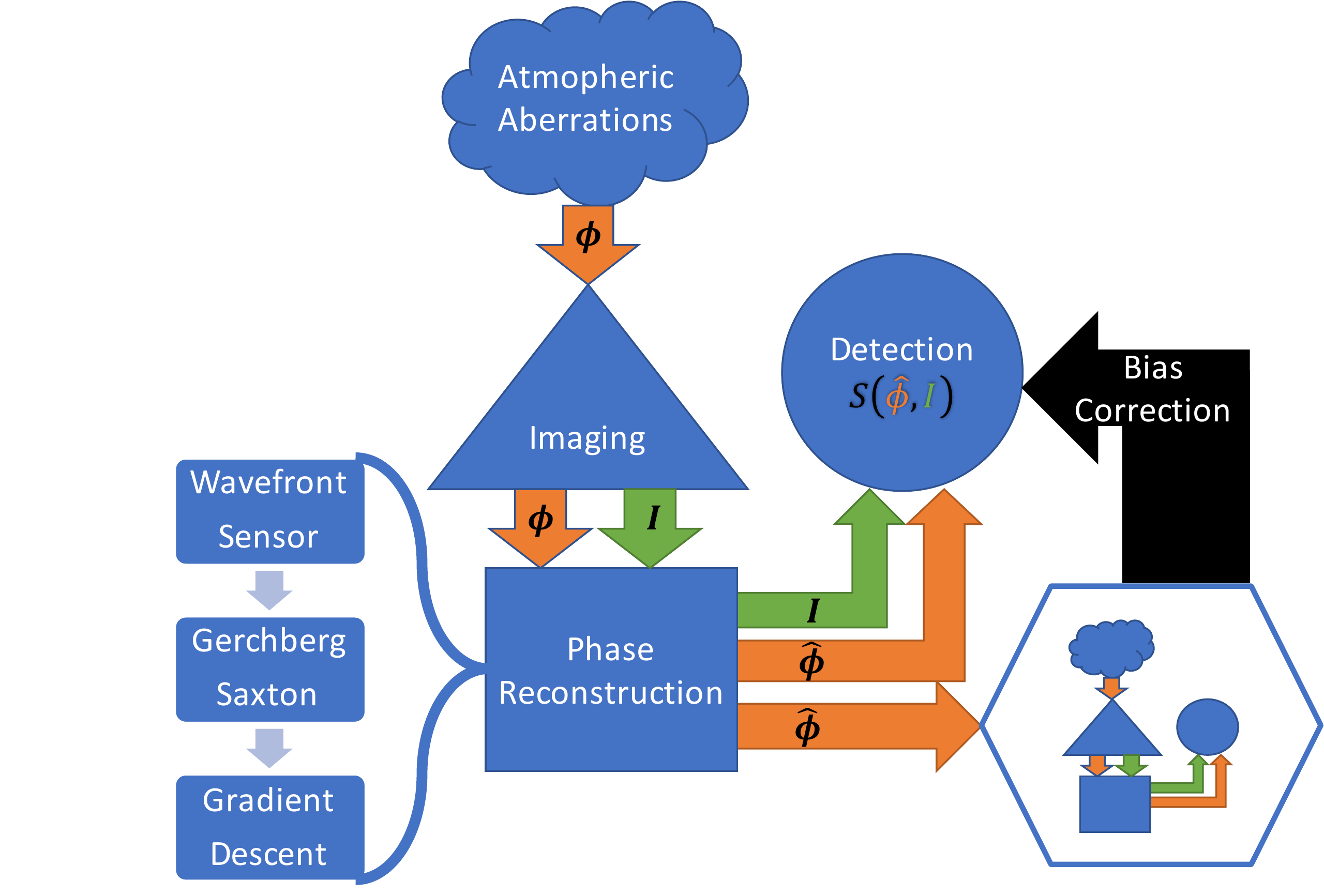}
\caption{Schematic diagram of the HCI method. Schematically in our method, we take short exposure images, reconstruct the phase aberrations for each image, calculate the statistic using the image and reconstructed phase and sample to correct for its bias.
\label{fig:method_scheme}}
\end{figure}

Motivated by this computation, in this work, we propose a method to perform HCI using short exposure speckle images, our scheme is illustrated in Figure \ref{fig:method_scheme}, the measurements are both imaging and wavefront sensing, a sequence of algorithms is used to reconstruct the atmospheric phase aberrations from the measurements \ref{subsec:rec_atm}, the optimal detection statistic is then calculated \ref{subsec:detect_stat} and Parametric Bootstrap method is used to correct the bias arising in the procedure \ref{subsec:bias_correction}. Analysis of the expected performance is presented in \ref{subsec:detect_stat} and \ref{subsec:overfit} and the results presented in Figure \ref{fig:contrast_curve}.

\section{High contrast imaging through the atmosphere} \label{sec:hci_turbulence}

When imaging from the ground, the atmosphere presents a difficulty. As air flows turbulently, different regions contain air of different velocities and densities that cause spatial variations in the index of refraction. As light passes through this medium with an in-homogeneous index of refraction, optical path differences accumulate, and phase aberrations are presented on the pupil plane of our telescope. Phase aberrations degrade the PSF of our telescope, and the atmospheric phase aberrations change rapidly, the instantaneous degraded PSF is typically comprised of diffraction-limited speckles, called speckle image, and the time-averaged called the seeing disk.

\subsection{Atmospheric seeing}\label{subsec:atm_seeing}
The PSF is a characteristic of the optical system that can be calculated for monochromatic illumination as the Fourier transform of the aperture function ($B$) and the phase aberration ($\phi$)
\begin{equation}
\label{eq:psf_mono}
    P = \left| \mathcal{F}\left[B e^{i \phi}\right] \right|^2 \, .
\end{equation}
In the case of perfect imaging from a circular aperture, the PSF will take the shape of an Airy disk with an angular scale of the diffraction limit $\lambda/D$.
The phase aberrations caused by the atmosphere degrade the PSF significantly and are a topic of many theoretical and observational studies. As a model for the atmospheric phase aberrations, we will use the classical Kolmogorov phase aberration structure function with a typical scale $r_0$ called the Fried parameter
\begin{eqnarray}
\label{eq:Kol_structure_function}
    D_\phi \left(\Delta \vec{x} \right) &\coloneqq& \langle \left( \phi (\vec{r}) - \phi (\vec{r}+\vec{\Delta x}) \right)^2 \rangle\nonumber\\
    &=& 6.88 \left( \frac{|\Delta \vec{x}|}{r_0} \right)^{5/3} \, .
\end{eqnarray}

The Fried parameter can also be understood as the size of the telescope that transitions from being diffraction limited to seeing limited.

To model imaging in a finite wavelength band, we integrate the single wavelength PSFs incoherently and include the two leading effects that change the PSF as a function of the wavelength:
\begin{itemize}
    \item The diffraction scale increases $\lambda_0/D \to \lambda/D$
    \item The aberration in phase units decreases $\phi \to \phi \lambda_0/\lambda$
\end{itemize}

\begin{equation}
\label{eq:psf_band}
    P = \int_{\lambda_l}^{\lambda_h} E_\lambda\left| \mathcal{S}_{\lambda/\lambda_0}\mathcal{F}\left[B e^{i \phi \frac{\lambda_0}{\lambda}}\right] \right|^2 d\lambda \, ,
\end{equation}
where $E_\lambda$ is a term representing the spectrum of the source, and $\mathcal{S}_{\lambda/\lambda_0}$ an operator that stretches the coordinates as the diffraction scale increases.

\subsection{Statistical model}\label{subsec:stat_model}
To push the limits of direct imaging we have to stay as close as possible to optimality, therefore we will start by mathematically formulating the high contrast imaging detection problem rigorously.

Directly imaging exoplanets can be formulated as a hypothesis testing question in which we want to distinguish between the null hypothesis of the images made of a single star with some flux $f$
\begin{equation}
\label{H0}
    H_0: \, T=f\delta_0 \, ,
\end{equation}
and the alternative hypothesis of an additional source with flux fraction $\epsilon$ at relative position $\vec{q}$
\begin{equation}
\label{H1}
    H_1: \, T=f\left((1-\epsilon)\delta_0 + \epsilon\delta_{\vec{q}}\right) \, .
\end{equation}

The mapping between the true sky $T$ and the measured images is by the PSF, with additive noise we will assume has a shot-noise component related to the source, shot-noise related to the sky background and detector read noise, taking the Gaussian approximation of the Poisson distribution (justified as we have a high number of photons per pixel)
\begin{equation}
    I \sim \mathcal{N}\left( P \otimes T ,  P \otimes T + b^2\right) \, ,
\end{equation}
where $\mathcal{N}$ is the normal distribution, $b$ is the noise from sky background and read-noise, $\otimes$ is the convolution operator.

\subsection{Detection statistic} \label{subsec:detect_stat}
To distinguish between the two hypotheses we will use the following statistic (with further discussion in Appendix \ref{app:stat_approx})
\begin{eqnarray}
\label{eq:statistic}
    S[q] &=& \left(\overleftarrow{P_{\hat{\phi}}} \otimes \frac{ I - f P_{\hat{\phi}}}{f P_{\hat{\phi}} + b^2} \right)_{\vec{q}} \nonumber\\
    &-& \left(\overleftarrow{P_{\hat{\phi}}} \otimes \frac{ I - f P_{\hat{\phi}}}{f P_{\hat{\phi}} + b^2} \right)_{0} \, ; \nonumber \\
    \hat{\phi} &=& \mathrm{argmax} \{P\left(I|H_0, P_{\phi}\right)P(\phi)\} \,
\end{eqnarray}
where $\overleftarrow{P}$ is the reverse of $P$, and ${\hat{\phi}}$ is the Maximum-A-Posteriori estimator of the atmospheric phase aberrations $phi$. We can read this statistic as a filter matching the image's deviation from a point source, and subtracting the location of the primary source as the unknown flux would create a flux deficit at its location. And the variance associated with the image's variance is
\begin{eqnarray}
\label{eq:statistic_var_opt}
    V_S = &f^2& \Bigg(\overleftarrow{P^2_{\hat{\phi}}}\otimes\frac{1}{f P_{\hat{\phi}} + b^2} \nonumber\\
    &-& 2\overleftarrow{P^2_{\hat{\phi}}}\otimes\frac{P_{\hat{\phi}}}{f P_{\hat{\phi}} + b^2}
    + \left[\overleftarrow{P^2_{\hat{\phi}}}\otimes\frac{1}{f P_{\hat{\phi}} + b^2}\right]_0\Bigg) \, .
\end{eqnarray}

We can estimate the $5\sigma$ detection threshold of this statistic for the ideal case in which we know $\phi$ exactly
\begin{equation}
    \min{\epsilon} \, : \, 5 \le \frac{E\left[S|H_1\right]}{\sqrt{E\left[V_S\right]}} \, ,
\end{equation}
by simulating atmospheric phase screens from the power spectrum calculated by \cite{noll_zernike_1976}, the resulting contrast thresholds as a function of separation for different bandwidths are shown in the colored dashed lines in Figure \ref{fig:contrast_curve}.

\section{Simulations} \label{sec:simulation}
To test our method we used end-to-end numerical simulations of the proposed measurement instruments.

We generate atmospheric aberrations by sampling the Fourier modes of the Kolmogorov spectrum as derived by \cite{noll_zernike_1976}. Images are calculated with the assumed true sky image and PSF as calculated in Equation \ref{eq:psf_band}, and then sampled according to Poisson with additional Gaussian noise. The simulation of the wavefront sensor neglects chromatic behavior, as only the centroid position will be used.

Reference values for simulation parameters are listed in Table \ref{tab:param_simul}.
\begin{table}[h]
    \centering
    \begin{tabular}{c | c }
        Parameter & Value\\
        \hline
        Pixel grid & $256\times256$\\
        Aperture diameter (physical) & $2m$\\
        Aperture diameter (pixels) & $85\frac{1}{3}px$\\
        Nyquist oversampling ratio & $1.5$\\
        Fried parameter & $15cm$\\
        Flux (per image, finite bandwidth) & $10^7$ @ $500-700nm$\\
        Background/read noise & 1 photon per pixel
    \end{tabular}
    \caption{Reference parameters for simulations.}
    \label{tab:param_simul}
\end{table}

\section{Methods} \label{sec:methods}
In order to use this statistic, we first need to show a method to calculate $\hat{\phi}$, and even though we know how the statistic distributes given the correct $\phi$, we need to examine carefully how it distributes when using $\hat{\phi}$.

\subsection{Recovering the atmospheric phase aberrations} \label{subsec:rec_atm}
In order to calculate the statistic from Equation \ref{eq:statistic} we have to find $\mathrm{argmax} \{P\left(I|H_0, P_{\phi}\right)P(\phi)\}$ , the Maximum-a-Posteriori (MAP) estimator for the atmospheric phase aberrations.

The general problem of phase retrieval, recovering a phase from its Fourier modulus in the presence of noise is known to be hard and a topic of many studies (for example the reviews \cite{fienup_phase_2013}, \cite{shechtman_phase_2015}). Therefore we employ a simple yet powerful sequence of algorithms.
Starting from direct measurement of the phase using a Shack-Hartmann WaveFront Sensor (SHWFS, \cite{platt_history_2001}), then improving that estimator using the Gerchberg-Saxton (GS, \cite{gerchberg_practical_1972}) algorithm and finally using Gradient Descent to ensure we get the MAP.

This procedure converges to the correct MAP for most instances in the case of enough photons ($\sim10^6$ per image), reasonable atmosphere ($r_0=15\mathrm{cm}@D=2\mathrm{m}$), but only for imaging in a single wavelength, further discussion, and elaborate results are presented in Appendix \ref{app:phase_rec}.

\subsection{Detection in the presence of learned PSF - $H_0$} \label{subsec:bias_correction}
When applying the statistic from Equation \ref{eq:statistic} on \emph{learned PSF} we have to deal with the effect of its inevitable errors. From simulations we learn, as expected, that our phase estimator can be modeled as distributing normally with some bias $\mu$ and covariance $\Sigma$ around the correct atmospheric phase aberration
\begin{equation}
\label{eq:estimator_distribution}
    \hat{\phi}\sim\mathcal{N}\left(\phi+\mu,\Sigma\right) \, .
\end{equation}
From Equations \ref{eq:psf_band} and \ref{eq:statistic} we can see that our statistic is a non-linear function of the phase estimator. Therefore, our estimator for the statistic, calculated using our estimator of the phase, is slightly \emph{biased}
\begin{equation}
\label{eq:bias_definition}
    \Delta S_\phi\equiv E\left[S_{\hat{\phi}} - S_\phi\right] \neq 0 \, ,
\end{equation}
but as it is a bias shared by all images, when we accumulate the statistic over many images it accumulates too and put an upper limit on the number of images we can use once it becomes significant.

As the notation suggests, this bias depends on $\phi$, and for simplicity, we will assume it varies slowly relative to our phase estimation error and therefore can be expanded in a series
\begin{eqnarray}
\label{eq:bias_expansion}
    \Delta S_{\hat{\phi}}[q] = \Delta S_{\phi}[q] &+& \frac{d\Delta S_{\phi}[q]}{d\phi}\Delta\phi \nonumber\\
    &+& \frac{1}{2}\Delta\phi\frac{d^2\Delta S_{\phi}[q]}{d\phi^2}\Delta\phi \, ,
\end{eqnarray}
and its expectation value
\begin{eqnarray}
\label{eq:bias_expectation}
    E_{\hat{\phi}}\left[\Delta S_{\hat{\phi}}[q]\right] = \Delta S_{\phi}[q] &+& \frac{d\Delta S_{\phi}[q]}{d\phi}\mu \nonumber\\
    &+& \frac{1}{2}\mu\frac{d^2\Delta S_{\phi}[q]}{d\phi^2}\mu\nonumber\\
    &+& \frac{1}{2}\mathrm{tr}\left(\frac{d^2\Delta S_{\phi}[q]}{d\phi^2}\Sigma\right) \, .
\end{eqnarray}

We extend the method of Parametric Bootstrap \cite{dekking_modern_2005} to estimate this bias. By sampling images and solving their corresponding phases based on the recovered phase estimator we can sample the distribution in Equation \ref{eq:estimator_distribution}
\begin{equation}
    \hat{\hat{\phi}}_1, \hat{\hat{\phi}}_2 \sim \mathcal{N}\left(\hat{\phi}+\mu,\Sigma\right) \, ,
\end{equation}
and calculate the following linear combination that has the expectation value as the bias we want to correct (detailed calculation in Appendix \ref{app:second_bias})
\begin{equation}
\label{eq:bias_correction}
E\left[\Delta S_{2\hat{\phi} - \hat{\hat{\phi}}_1} - \Delta S_{2\hat{\hat{\phi}}_1 - \hat{\phi}} + \Delta S_{\hat{\hat{\phi}}_1 + \hat{\hat{\phi}}_2 - \hat{\phi}}\right] = \Delta S_\phi \, .
\end{equation}

\begin{figure}[htbp]
\gridline{\fig{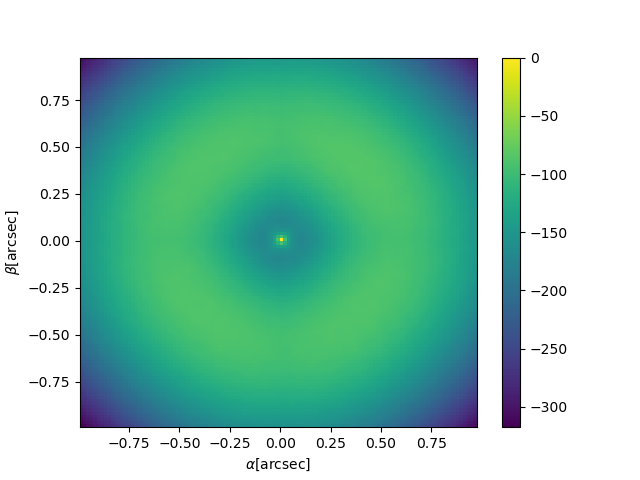}{1.2\linewidth}{(a)\label{subfig:biased_stat}}}
\gridline{\fig{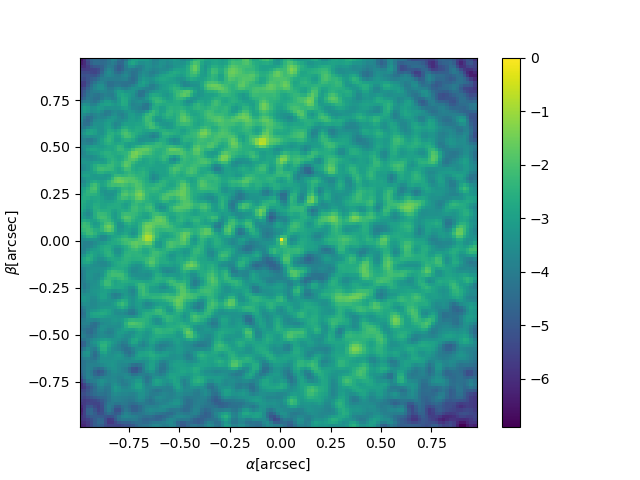}{1.2\linewidth}{(b)\label{subfig:debiased_stat}}}
\caption{Detection statistic (a) without and (b) with bias correction. The images were simulated for a $2\mathrm{m}$-telescope, at seeing condition $r_0=15\mathrm{cm}$, for a bandwidth of 10\%, and the statistic was accumulated over $10^4$ images.
\label{fig:stat_debias}}
\end{figure}

In practice, we calculate this term a few times to get better convergence to the mean and have a handle on its variance, which is empirically smaller than the inherent variance of the statistic, that is calculated in Equation \ref{eq:statistic_var_opt}, and can be reduced as $1/\sqrt{N}$ with more simulations.

\subsection{Signal loss due to over fitting} \label{subsec:overfit}
After we made sure we are recovering null detections for images sampled from $H_0$, we need to test the expected performance of the method for signals from $H_1$.

An inevitable tension is present in our phase aberration estimation process. The posterior probability of the phase aberration, $\phi$, is dominated by the deviation of its corresponding PSF, $P_\phi$, from the image. This will lead, in the presence of an additional source, to \emph{overfitting} of the secondary as part of the primary.
Detecting using an overfitted $\hat{\phi}$ subtracts part of the secondary and therefore loses some of the signal available with perfect knowledge of $\phi$. To quantify this effect, we perform injection-recovery simulations for different separations and different brightness.

\begin{figure}[htbp]
\gridline{\fig{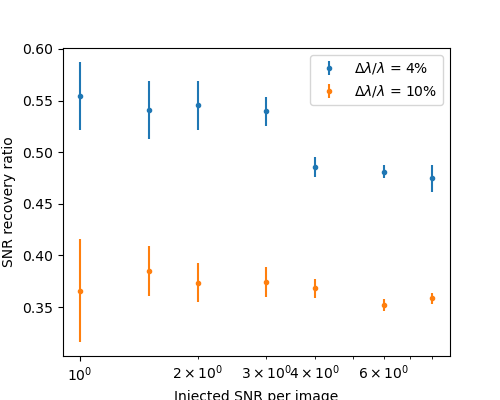}{1.1\linewidth}{(a)\label{fig:inj_rec_snr}}}
\gridline{\fig{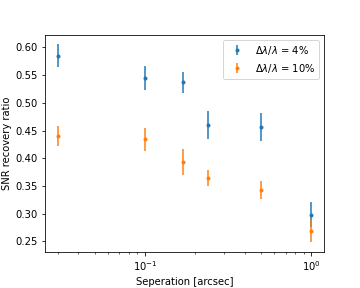}{1.1\linewidth}{(b)\label{fig:inj_rec_sep}}}
\caption{Injection-Recovery simulation. (a) The recovered ratio for different injected SNR per image, injection at $|q|=0.2^{\prime\prime}$, we can see a plateau as we go for weaker signals, (b) recovery ratio for different separations at $\mathrm{SNR_{inj}=1}$ per image.
\label{fig:inj_rec}}
\end{figure}

In general, we expect the recovered SNR to be some function of the injected SNR and the separation from the star at which the signal is injected. In the weak signal limit at which we work, the injected SNR per image is small $\mathrm{SNR_{inj}} \ll 1$, and we expect the recovered SNR to be proportional to the injected SNR, as can be seen in Figure \ref{fig:inj_rec}a, and the dependence on the location to be some smooth function that takes into account the assumed phase aberrations spectrum and the degeneracy between the phase aberrations and the additional source, as can be seen in Figure \ref{fig:inj_rec}b.
\begin{equation}
    \mathrm{SNR_{inj}} = \alpha(|q|) \mathrm{SNR_{inj}} \, .
\end{equation}

We use the values of $\alpha(|q|)$ from Figure \ref{fig:inj_rec} to convert the optimal dotted lines in Figure \ref{fig:contrast_curve} to the more realistic solid lines.
In the future, we will investigate using several wavelength bands in tandem to mitigate this loss, using the a-chromatic position of the secondary and the chromatic effect of the phase aberrations.

\section{Conclusion and outlook}\label{sec:conclusions}
We presented a method to reconstruct the atmospheric phase aberrations using phase retrieval algorithms, and use its corresponding PSFs to calculate the optimal statistic to detect the presence of a secondary faint source in a sequence of short-exposure speckle images. We presented a method to correct the arising bias in our statistic, showing its complete correction for stacks of up to $10^4$ images, and a pathway to increase it by expanding Equation \ref{eq:bias_expansion} beyond second order, which is crucial for the use of the detection method.

The leading idea for our method is to optimize the PSF of the image through parameters of a physically constrained model, use the optimal statistic, and correct for its arising bias or other artifacts. The analysis presented in this work treats the case of imaging with no AO, but the method isn't limited to such observations. A probable avenue for improvement is to use AO to some extent which will reduce the photon noise of the host star, and concentrate the light of the planet to a diffraction-limited spot, thus enabling even better detection of the secondary source.

To advance this method to achieve high contrast on sky, further work is required, including:
\begin{itemize}
    \item Better phase retrieval scheme needs to be devised, to enable the retrieval from images taken in a narrow wavelength band.
    \item Demonstration of the method in-lab is essential and will ensure we can precisely describe speckle images.
    \item Generalizing the method for simultaneous imaging in a few bands, which is important for collecting enough light, and might make the phase retrieval more regular and therefore easier.
\end{itemize}

\section*{Acknowledgements}
BZ is supported by a research grant from the Center for New Scientists at the Weizmann Institute of Science and a research grant from the Ruth and Herman Albert Scholarship Program for New Scientists.
BZ and DG are supported by the Israeli Council for Higher Education: Competitive Program for Data Science \& AI Research Centers.

\appendix

\section{Approximated Statistic} \label{app:stat_approx}
As shown by \cite{neyman_problem_1933}, when testing hypotheses the likelihood ratio test achieves maximal detection power for a given false positive probability, therefore we will employ it for our detection problem
\begin{equation}
\label{eq:likelihood_ratio}
    \Lambda\left(I\right) = \frac{P\left(I|H_1\right)}{P\left(I|H_0\right)} \, ,
\end{equation}
which we can marginalize over the atmospheric aberrations
\begin{equation}
\label{eq:marginilised_likelihood_ratio}
\Lambda\left(I\right) = \frac{\int P\left(I|H_1, P_{\phi}\right)P(\phi)d\phi}{\int P\left(I|H_0, P_{\phi}\right)P(\phi)d\phi} \, .
\end{equation}

Calculating explicitly both marginalized likelihoods is impractical so we employ the Gaussian approximation. At the limit of well-measured phase aberrations, in which we are interested, the posterior distribution of the phase aberrations is well approximated as a Gaussian.

Under the Gaussian approximation, the integrals reduce to the value at maximum times the square root of its covariance determinant.
We further approximate the MAP estimate under $H_1$ as the MAP estimate under $H_0$. This approximation is justified as in a single image we cannot favor $H_1$, so specifically, we cannot detect a difference in the MAP.

The covariance also doesn't contribute, as it presents a logarithmic correction while being nearly identical for the null and alternative hypotheses.

\begin{eqnarray}
\label{eq:marginilised_likelihood_ratio_map}
\Lambda\left(I\right) &\approx& \frac{P\left(I|H_1, P_{\hat{\phi}}\right)P(\hat{\phi})}{P\left(I|H_0, P_{\hat{\phi}}\right)P(\hat{\phi})} \, ; \nonumber\\
\hat{\phi} &=& \mathrm{argmax} \{P\left(I|H_0, P_{\phi}\right)P(\phi)\} \, .
\end{eqnarray}

and for the case of a companion detection with known PSF, as we assume here, the likelihood-ratio test can be simplified to the expression in Equation \ref{eq:statistic}, as calculated by \cite{nir_possible_2019}.

\section{Phase recovery procedure} \label{app:phase_rec}
\subsection{Shack-Hartmann Wavefront Sensor} \label{subapp:shwfs}

The SHWFS is a device that images small areas of the aperture separately. For sufficiently small areas (of the order of $r_0$) the atmospheric phase aberrations can be estimated as simply linear. When imaging a wavefront with some constant slope the image is an off-centered point, with the tip-tilt of the point related linearly to the slope of the wavefront.

In our simulation, we used an array of $26\times26$ sub-apertures which for our reference seeing conditions and telescope size resulted in each sub-aperture being of the size $\sim r_0/2\times r_0/2$, to recover the phase aberrations from the WFS we used a maximum-a-posteriori estimator in the Gaussian regime and using a measured design matrix, as done by \cite{clare_wavefront_2004}.

\begin{figure}[htbp]
\gridline{\fig{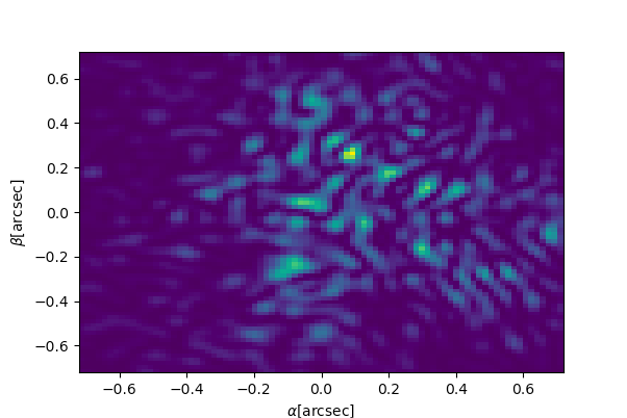}{0.4\linewidth}{(a)\label{fig:wfs_in}}
\fig{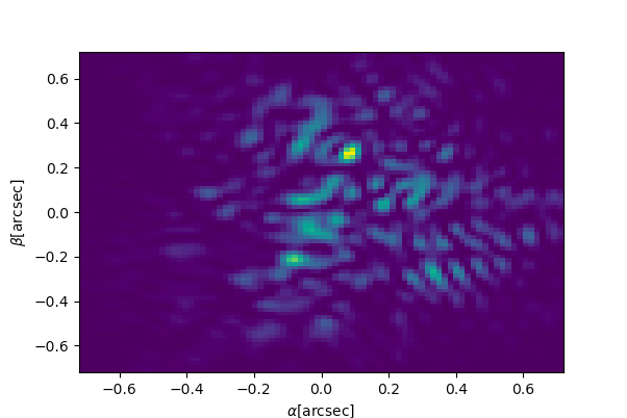}{0.4\linewidth}{(b)\label{fig:wfs_out}}}
\caption{Comparison between a (a) sampled PSF and the (b) the PSF corresponding to the phase measured by the SHWFS. An obvious, order-unity difference is visible between the two PSFs, therefore the measured phase is not accurate enough to be used for detection.
\label{fig:wfs_psf}}
\end{figure}

The phase aberrations recovered by the WFS have significant errors, with the PSF calculated using them having order-unity deviations from the input PSF as can be seen in Figure \ref{fig:wfs_psf}, therefore cannot be used to calculate the statistic. To improve the phase estimator we use further optimization techniques.

\subsection{Gerchberg-Saxton algorithm} \label{subapp:gs}
The Gerchberg-Saxton algorithm is an algorithm that was designed and proved to solve the phase retrieval problem under the $l_2$ norm by \cite{gerchberg_practical_1972}. We started the algorithm with the initial guess of the pupil function and the phase measured by the WFS, from that the algorithm transforms the guessed field to the image plane, constraining the amplitude by the measured image, then returning to the pupil plane constraining the pupil function again and repeating until convergence or some maximal number of iterations.

To quantify the convergence of the algorithm we compare an observable, likelihood-based criteria
\begin{equation}
    \sum \frac{\left(I-fP\right)^2}{fP+b^2} \lessgtr N_{\text{pixel}} \, ,
\end{equation}
and an unobservable, phase error criteria, we simulated images in the single wavelength case for different seeing and flux conditions we examine the performance of the algorithm in the observable criteria and its agreement with the unobservable one.

\newcommand*{\MinNumberA}{50}%
\newcommand*{\MaxNumberA}{100}%

\newcommand{\ApplyGradientA}[1]{%
    \pgfmathsetmacro{\PercentColor}{100.0*(#1-\MinNumberA)/(\MaxNumberA-\MinNumberA)}
    \textcolor{green!\PercentColor!red}{#1}
}
\begin{table}[h]
    \centering
    \begin{tabular}{c | c c c c}
        \diagbox{$r_0[\mathrm{cm}]$}{$N_{\mathrm{photons}}$} & $10^5$ & $3\times10^5$ & $10^6$ & $3\times10^6$ \\ \hline
        15 & \ApplyGradientA{100} & \ApplyGradientA{90}   & \ApplyGradientA{67}  & \ApplyGradientA{65}  \\
        18 & \ApplyGradientA{100} & \ApplyGradientA{98}   & \ApplyGradientA{93}  & \ApplyGradientA{89}  \\
        21 & \ApplyGradientA{100} & \ApplyGradientA{100}  & \ApplyGradientA{98}  & \ApplyGradientA{96}  \\
        24 & \ApplyGradientA{100} & \ApplyGradientA{100}  & \ApplyGradientA{100} & \ApplyGradientA{100} \\
    \end{tabular}
    \caption{Success rate percent for different seeing conditions and photons per speckle image (in the used sub-band). The success rate was determined according to the likelihood threshold for $10^3$ attempts.}
    \label{tab:gs_success}
\end{table}

In Table \ref{tab:gs_success} we show the fraction of phase aberrations that are successful in converging to a solution that passes the likelihood threshold, we see that the algorithm converges more for smaller phase aberrations or flux. However, in Table \ref{tab:gs_phase} we see the phase error of runs that converged successfully in the likelihood sense is high for the small flux case, which we can understand as not constraining enough. To ensure the gradient-based method will work we want the phase error to be smaller than a radian so the problem will be close to quadratic or some low-order polynomial.

\newcommand*{\MinNumberB}{0}%
\newcommand*{\MaxNumberB}{1}%

\newcommand{\ApplyGradientB}[1]{%
    \pgfmathsetmacro{\PercentColor}{100.0*(#1-\MinNumberB)/(\MaxNumberB-\MinNumberB)}
    \textcolor{red!\PercentColor!green}{#1}
}

\begin{table}[h]
    \centering
    \begin{tabular}{c | c c c c}
        \diagbox{$r_0[\mathrm{cm}]$}{$N_{\mathrm{photons}}$} & $10^5$ & $3\times10^5$ & $10^6$ & $3\times10^6$ \\ \hline
        15 & \ApplyGradientB{0.79} & \ApplyGradientB{0.45}   & \ApplyGradientB{0.16}  & \ApplyGradientB{0.05}  \\
        18 & \ApplyGradientB{0.81} & \ApplyGradientB{0.43}   & \ApplyGradientB{0.18}  & \ApplyGradientB{0.06}  \\
        21 & \ApplyGradientB{0.71} & \ApplyGradientB{0.41}   & \ApplyGradientB{0.19}  & \ApplyGradientB{0.07}  \\
        24 & \ApplyGradientB{0.80} & \ApplyGradientB{0.40}   & \ApplyGradientB{0.19}  & \ApplyGradientB{0.08}  \\
    \end{tabular}
    \caption{Phase root-mean-square for different seeing conditions and photons per image. The phase root-mean-square was averaged for runs that passed the likelihood threshold.}
    \label{tab:gs_phase}
\end{table}

In future works we will expand this optimization step to allow convergence for finite bandwidth imaging.

\subsection{Gradient descent} \label{subapp:grad}
To go from the $l_2$ solution to the MAP estimator we optimize a function proportional to the log-posterior

\begin{equation}
    \mathcal{L}(\phi) \coloneqq \sum_{x,y} \frac{\left( I - f P_\phi \right)^2}{f P_\phi + b^2} + \phi^T C^{-1} \phi \, ,
    \label{eq:cost_func}
\end{equation}

using gradient descent with the method to rapidly calculate the gradient as done by \cite{fienup_phase_1999}, 
this method takes advantage of the Fourier transform in Equation \ref{eq:psf_band} to calculate it with only 2 FFT operations

\begin{equation}
    \frac{\partial \mathcal{L}}{\partial \phi_u} 
    = \left[2\phi^T C^{-1}\right]_u + \int_{\lambda_l}^{\lambda_h} \Re{\left\{ -i B \frac{\lambda_0}{\lambda} e^{-i \phi \frac{\lambda_0}{\lambda}} \mathcal{F}^{-1}\left[ 2f\left(\mathcal{S}^{-1}_{\lambda/\lambda_0}\frac{\partial \mathcal{L}_I}{\partial P}\right)  \mathcal{F}\left[B e^{i \phi \frac{\lambda_0}{\lambda}}\right]\right] \right\}_u } d\lambda \, .
    \label{eq:cost_gradient}
\end{equation}

\section{Second order bias estimate} \label{app:second_bias}
To prove Equation \ref{eq:bias_correction} we start from $\hat{\hat{\phi}}_1$ and $\hat{\hat{\phi}}_2$ that are estimates for $\hat{\phi}$ and i.i.d according to $\mathcal{N}\left(\hat{\phi} + \mu, \Sigma\right)$, examining each term separately, first an unbiased phase estimate
\begin{eqnarray}
2\hat{\phi} - \hat{\hat{\phi}}_1 &=& \hat{\phi} + (\hat{\phi} - \hat{\hat{\phi}}_1)\nonumber\\
&\sim& \mathcal{N}\left(\phi, 2\Sigma\right)\nonumber\\
E_{\hat{\phi}, \hat{\hat{\phi}}_1}\left[\Delta S(2\hat{\phi} - \hat{\hat{\phi}}_1)\right] &=& 
\Delta S(\phi) + \mathrm{tr}\left(\Sigma \frac{\partial^2 \Delta S}{\partial \phi^2}|_{\phi}\right) \, ,
\end{eqnarray}
and two biased phase estimates with different variance
\begin{eqnarray}
2\hat{\hat{\phi}}_1 - \hat{\phi} &=& 2(\hat{\hat{\phi}}_1 - \hat{\phi}) + \hat{\phi} \nonumber\\
&\sim& \mathcal{N}\left(\phi + 3\mu, 5\Sigma\right) \nonumber\\
E_{\hat{\phi}, \hat{\hat{\phi}}_1}\left[\Delta S(2\hat{\hat{\phi}}_1 - \hat{\phi})\right] &=& 
\Delta S(\phi) + 
3\mu \frac{\partial\Delta S}{\partial \phi}|_{\phi} +
\frac{9}{2}\mu \frac{\partial^2 \Delta S}{\partial \phi^2}|_{\phi} \mu +
\frac{5}{2}\mathrm{tr}\left(\Sigma \frac{\partial^2 \Delta S}{\partial \phi^2}|_{\phi}\right) \, ,
\end{eqnarray}
\begin{eqnarray}
\hat{\hat{\phi}}_1 + \hat{\hat{\phi}}_2 - \hat{\phi} &=& (\hat{\hat{\phi}}_1 - \hat{\phi}) + (\hat{\hat{\phi}}_2 - \hat{\phi}) + \hat{\phi} \nonumber\\
&\sim& \mathcal{N}\left(\phi + 3\mu, 3\Sigma\right) \nonumber\\
E_{\hat{\phi}, \hat{\hat{\phi}}_1, \hat{\hat{\phi}}_2}\left[\Delta S(\hat{\hat{\phi}}_1 + \hat{\hat{\phi}}_2 - \hat{\phi})\right] &=&
\Delta S(\phi) + 
3\mu \frac{\partial\Delta S}{\partial \phi}|_{\phi} +
\frac{9}{2}\mu \frac{\partial^2 \Delta S}{\partial \phi^2}|_{\phi} \mu +
\frac{3}{2}\mathrm{tr}\left(\Sigma \frac{\partial^2 \Delta S}{\partial \phi^2}|_{\phi}\right) \, .
\end{eqnarray}
Inserting those relations to the LHS of Equation \ref{eq:bias_correction} easily leads to the RHS.

\bibliography{main}{}
\bibliographystyle{aasjournal}



\end{document}